\pdfoutput=1
\documentclass[%
showpacs,
 amsmath,amssymb,
 aps,
 superscriptaddress,
 pra,
floatfix,
twocolumn, %
]{revtex4-1}

\usepackage{graphicx}
\usepackage{dcolumn}
\usepackage{bm}
\usepackage[utf8]{inputenc}
\usepackage{amsmath}
\usepackage[amssymb]{SIunits}
\usepackage{amsfonts}
\usepackage{amssymb}
\usepackage{mathtools}
\usepackage{dsfont}
\usepackage{color}
\usepackage{isotope}
\usepackage{calc}
\usepackage{braket}

\begin{document}


\title{Spontaneous Emission from a Two-Level Atom in a Rectangular Waveguide} %

\author{Moochan B.~Kim}
\affiliation{Hearne Institute for Theoretical Physics %
and Department of Physics and Astronomy, %
Louisiana State University, %
Baton Rouge, %
Louisiana 70803, %
USA} %

\author{Georgios Veronis}
\affiliation{Division of Electrical and Computer Engineering, Louisiana State University, Baton Rouge, Louisiana 70803, USA} %
\affiliation{Center for Computation and Technology, Louisiana State University, Baton Rouge, Louisiana 70803, USA} %

\author{Tae-Woo Lee}
\affiliation{Center for Computation and Technology, Louisiana State University, Baton Rouge, Louisiana 70803, USA} %

\author{Hwang Lee}
\affiliation{Hearne Institute for Theoretical Physics %
and Department of Physics and Astronomy, %
Louisiana State University, %
Baton Rouge, %
Louisiana 70803, %
USA} %

\author{Jonathan P.~Dowling%
}%
\affiliation{Hearne Institute for Theoretical Physics %
and Department of Physics and Astronomy, %
Louisiana State University, %
Baton Rouge, %
Louisiana 70803, %
USA} %
\affiliation{Computational Science Research Center, Beijing, %
100084, China} %

\date{\today}%

\begin{abstract} %
Quantum mechanical treatment of light inside dielectric media is important to understand the behavior of an optical system. %
In this paper, %
a two-level atom embedded in a rectangular waveguide surrounded by a perfect electric conductor is considered. %
Spontaneous emission, propagation, and detection of a photon are described by the second quantization formalism. %
The quantized modes for light are divided into two types: photonic propagating modes %
and localized modes with exponential decay along the direction of waveguide. 
Though spontaneous emission depends on all possible modes including the localized modes, %
detection 
far from the source only depends on the propagating modes. %
This discrepancy of dynamical behaviors gives two different decay rates along space and time %
in the correlation function of the photon detection. 
\end{abstract} %


\maketitle

\section{Introduction} %

The photon is one of the fascinating objects that has been researched since Einstein's introduction. %
Usually, the currently accepted definition of photon is a monochromatic Fourier mode for electromagnetic waves %
in the vacuum~\cite{NatureLight2008}. %
Since arbitrary shapes of photonic modes can be generated by superposition of modes, 
this usually accepted definition has been successfully used %
to describe many different physical processes~\cite{ScullZubairy1997}. %

Recently, the technology of dielectric material fabrication is able to handle 
small sample sizes, %
comparable to the wavelengths of light~\cite{Faraon2008,Dowling2003}. %
Generation of a single photon inside the dielectric media and propagation through waveguide %
has been considered~\cite{Englund2007}. %
In that investigation, a creation (annihilation) operator of the photon was used, %
which propagates to the right or left. %
Since this system is a one-dimensional waveguide, %
$e^{ikx}$-type of modes are enough to handle problems for the generation and propagation of photons. %
Although an integrated photonic crystal circuit %
operates at the single-photon level %
and is approximated as one-dimensional system, %
the actual phenomena occur in three dimensions. %
So, we need to treat the problem quantum-mechanically with Maxwell's equations in three dimensions, %
or the second quantization method to describe photon's behavior inside the dielectric material. %
Furthermore, from several formalisms~\cite{Philbin2010}, %
we also need to discern a proper description for photon inside dielectric material %
by comparing experimentally verifiable quantities. %

To initiate such investigation, this paper has considered a simple model system %
to treat the generation, propagation, and detection of light inside a dielectric material. %
We have found that the spontaneous decay can induce exponentially localized modes, %
which are usually ignored in the propagation problem, %
and that there are two different decay rates for the space and time in their correlation function. 
Though the exponentially localized modes satisfy the transversality condition, $\nabla \cdot \mathbf{E} = 0$, %
they do not propagate~\cite{CohenTannoudji1989}. %
So only the nonlocalized modes are propagating modes, %
which are present in the far-field region. %
However, in near field region the photon detector measures the whole electric field, %
and the distinction of propagating modes from the localized modes is not possible through photon detection. %

The remainder of the paper is organized as follows. %
In Sec.~II we derive the electric field operator using the second quantization method %
according to Glauber's approach~\cite{Glauber1991}. %
In Sec.~III, we treat spontaneous emission in a two-level atom %
using the electric field operator derived in Sec.~II. %
Finally, we calculate the spatio-temporal correlation function, %
and in Sec.~IV we summarize our conclusions. %

\section{Electric Field Operator in Rectangular PEC waveguide} %

\begin{figure}[b] %
\centering %
\includegraphics[width=0.8\linewidth]{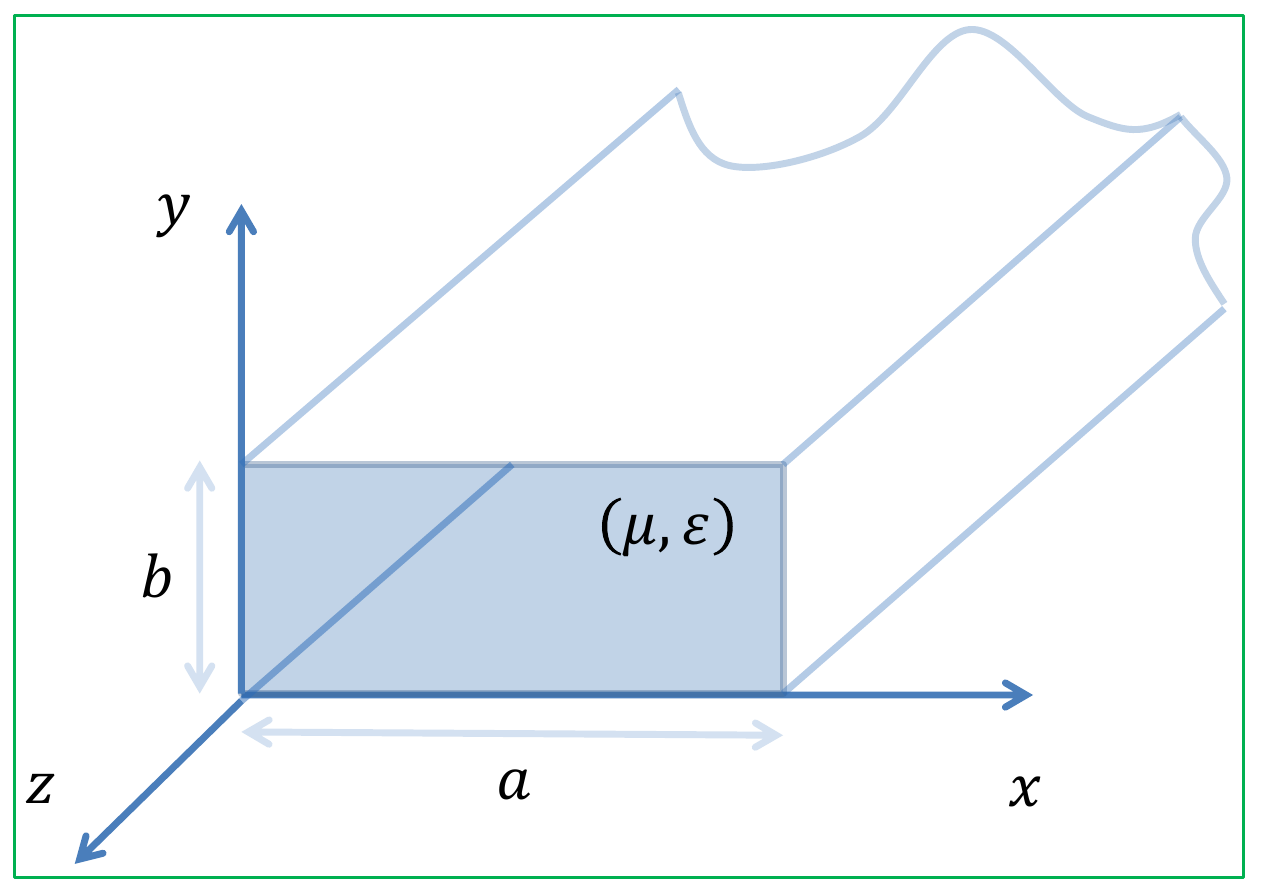} %
\caption{Diagram for a rectangular waveguide along $z$-axis. %
The waveguide is filled with a dielectric material with electric permittivity $\epsilon$ and magnetic permeability $\mu$. %
We assume that $a > b$. (Color online)} %
\label{Fig:RectangularWaveguide} %
\end{figure}

To see the effect of dielectric materials on photon emission, propagation, and detection, %
let's consider a rectangular waveguide filled by a dielectric material %
with constant electric permittivity $\epsilon$ and constant magnetic permeability $\mu$. %
Here, we choose a rectangular waveguide since it has a simple though nontrivial geometry. %
For simplicity, %
we consider that the dielectric material inside the waveguide is surrounded by a perfect electric conductor (PEC) %
so that the tangential electric field at boundary is zero. %
Let's put the waveguide along the $z$-axis, and the center of coordinates at the vertex of the rectangles %
with the length along $x$-axis as $a$ %
and that along $y$-axis as $b$ with $a > b$, as shown in FIG.~\ref{Fig:RectangularWaveguide}. %

The Maxwell equations in source free region are %
\begin{align} %
\nabla \cdot \mathbf{D} = 0, \\ %
\nabla \cdot \mathbf{B} = 0, \\ %
\label{Eq:AmpereMaxwell} %
\nabla \times \mathbf{E} = - \frac{\partial \mathbf{B}}{\partial t}, \\ %
\label{Eq:FaradayDisplacementMaxwell} %
\nabla \times \mathbf{H} = \frac{\partial \mathbf{D}}{\partial t}, %
\end{align} %
with constituent equations $\mathbf{D} = \epsilon \mathbf{E}$ and $\mathbf{B} = \mu \mathbf{H}$. %

To get the field operator using the second quantization method, %
we need complete eigenmodes from the classical Maxwell's wave equations~\cite{Glauber1991}. %
By assuming a harmonic time dependence $e^{-i\nu t}$ with frequency $\nu$, %
which is considered as eigenvalue, %
Eqs.~\eqref{Eq:AmpereMaxwell} and \eqref{Eq:FaradayDisplacementMaxwell} become %
\begin{align} %
\label{Eq:TimeIndependentAmpere} %
\nabla \times \mathbf{E} = i \nu \mathbf{B} = i \nu \mu \mathbf{H}, \\ %
\label{Eq:TimeIndependentFaradayDisplacementMaxwell} %
\nabla \times \mathbf{H} = -i \nu \mathbf{D} = -i\nu\epsilon\mathbf{E}. %
\end{align} %
The Maxwell's equations \eqref{Eq:TimeIndependentAmpere} and \eqref{Eq:TimeIndependentFaradayDisplacementMaxwell} %
lead to wave equations for $\mathbf{E}$ and $\mathbf{H}$: %
\begin{align}  %
\label{Eq:WaveEqnElectric} %
\nabla^2 \mathbf{E} + k^2 \mathbf{E} = 0, \\ %
\label{Eq:WaveEqnMagnetic} %
\nabla^2 \mathbf{H} + k^2 \mathbf{H} = 0,  %
\end{align} %
with $k = \nu \sqrt{\epsilon \mu}$. %
If we assume the propagation along $z$-axis as $e^{-\gamma z}$, %
the longitudinal field components ($E_x, E_y, H_x$ and $H_y$) can be written %
in terms of the transverse field components ($E_z$ and $H_z$) as follows: %
\begin{align} %
E_x & = \frac{1}{h^2} \left( -\gamma \frac{\partial E_z}{\partial x} - i \nu \mu \frac{\partial H_z}{\partial y} \right), \\ %
E_y & = \frac{1}{h^2} \left( -\gamma \frac{\partial E_z}{\partial y} + i \nu \mu \frac{\partial H_z}{\partial z} \right), \\ %
H_x & = \frac{1}{h^2} \left( i \nu \epsilon \frac{\partial E_z}{\partial y} - \gamma \frac{\partial H_z}{\partial x} \right), \\ %
H_y & = \frac{1}{h^2} \left( - i \nu \epsilon \frac{\partial E_z}{\partial x} - \gamma \frac{\partial H_z}{\partial y} \right) %
\end{align} %
with $h^2 = \gamma^2 + k^2$. %
Since $\gamma = \sqrt{h^2 - k^2}$, $\gamma$ will be pure imaginary for $\nu > \nu_c$ or pure real otherwise. %
Here, $\nu_c = h/\sqrt{\epsilon \mu}$ is the cutoff-frequency. %

The solutions of the wave-equation are divided into two types of modes: 
the transverse magnetic (TM) modes with $H_z = 0$, %
and the transverse electric (TE) modes with $E_z = 0$. %
The field components for the TM modes are %
\begin{align} %
E_z^{\text{TM}} & = E_0 \sin \frac{m\pi x}{a} \sin \frac{n\pi y}{b} e^{-\gamma z}, \\ %
E_x^{\text{TM}} & = - \frac{\gamma}{h^2} \left( \frac{m\pi}{a} \right) E_0 \cos\frac{m\pi x}{a} \sin\frac{n\pi y}{b} e^{-\gamma z}, \\ %
E_y^{\text{TM}} & = - \frac{\gamma}{h^2} \left( \frac{n\pi}{b} \right) E_0 \sin\frac{m\pi x}{a} \cos\frac{n\pi y}{b} e^{-\gamma z}, \\ %
H_x^{\text{TM}} & = \frac{i\nu\epsilon}{h^2}\left(\frac{n\pi}{b}\right) E_0 \sin\frac{m\pi x}{a} \cos\frac{n\pi y}{b} e^{-\gamma z}, \\ %
H_y^{\text{TM}} & = -\frac{i\nu\epsilon}{h^2}\left(\frac{m\pi}{a}\right) E_0 \cos\frac{m\pi x}{a} \sin\frac{n\pi y}{b} e^{-\gamma z}, %
\end{align} %
with positive integers $m$ and $n$,
and for TE modes are %
\begin{align} %
H_z^{\text{TE}} & = H_0 \cos \frac{m\pi x}{a} \cos \frac{n\pi y}{b} e^{-\gamma z}, \\ %
E_x^{\text{TE}} & = \frac{i\nu\mu}{h^2} \left(\frac{n\pi}{b}\right) H_0\cos \frac{m\pi x}{a} \sin \frac{n\pi y}{b} e^{-\gamma z}, \\ %
E_y^{\text{TE}} & = - \frac{i\nu\mu}{h^2} \left(\frac{m\pi}{a}\right) H_0\sin \frac{m\pi x}{a} \cos \frac{n\pi y}{b} e^{-\gamma z}, \\ %
H_x^{\text{TE}} & = \frac{\gamma}{h^2} \left(\frac{m\pi}{a}\right) H_0\sin \frac{m\pi x}{a} \cos \frac{n\pi y}{b} e^{-\gamma z}, \\ %
H_y^{\text{TE}} & = \frac{\gamma}{h^2} \left(\frac{n\pi}{b}\right) H_0\cos \frac{m\pi x}{a} \sin \frac{n\pi y}{b} e^{-\gamma z}, %
\end{align} %
with non-negative integers $m$ and $n$ without simultaneous $m= 0$ and $n=0$. %
The boundary conditions suggest that %
allowable $\gamma_{mn} = \sqrt{ h_{mn}^2 - k^2}$ %
with $h_{mn} = \sqrt{ \left( m\pi/a\right)^2 + \left( n\pi/b\right)^2}$. %
The propagating mode with the lowest cutoff frequency is a TE mode with $m = 1$ and $n = 0$. 

In the usual treatment of the propagation problem with injected light as input along the waveguide, %
the modes with exponential behavior ($\gamma > 0$) have been excluded %
since they diverge at $\vert z \vert \rightarrow \infty$, %
and only the oscillating eigenmodes ($\gamma = i\beta$) 
are considered, 
since they can propagate into the far-field region, where the detector is located. %
However, in our system the light is generated inside the dielectric material, %
and we want to solve a photon-generation problem in quantum mechanics. %
So, the usually ignored exponential decay modes can operate as one of the quantized modes, %
which can interact with the embedded atom. %
For a single atom located at $z=0$, %
the replacement of the exponential function $e^{-\gamma z}$ into $e^{-\gamma \vert z \vert}$ %
is an acceptable eigenmode with a singularity in first derivative at $z=0$. %
One thing to note is that all eigenmodes satisfy the transversaility condition, %
that is $\nabla \cdot \mathbf{E}^{\text{TE}(\text{TM})}_{mn} = 0$. %
Since the corresponding wave vector $\mathbf{k}$ for the localized modes %
has a complex number in component along the $z$-axis, %
these modes do not propagate and only oscillate in time near the source. %

The corresponding electric field operator can be written as~\cite{Glauber1991} %
\begin{align} %
\label{Eq:EFieldOperator} %
\hat{\mathbf{E}} & = \sum_{\nu} 
\sum_K %
\left\{ \mathbf{E}_K^\nu \hat{a}_K^\nu e^{-i\nu t} + (\mathbf{E}_K^\nu)^* (\hat{a}_K^\nu)^\dagger e^{i\nu t} \right\} 
\end{align} %
where $K = \{ \text{TE} (\text{ or } \text{TM}), m, n \}$ denotes the set of quantum numbers for eigenmodes, %
and $\hat{a}_K^\nu$ ( $\hat{a}_K^{\nu\dagger}$ ) is annihilation (creation) operator for the corresponding mode. %
$\mathbf{E}_K^\nu$ are the eigenmodes for the electric field for a fixed frequency %
with orthogonality %
\begin{align} %
\int d\mathbf{r} \, \epsilon \mathbf{E}_K^\nu \cdot ( \mathbf{E}_{K'}^\nu)^* = 0 \qquad \text{ for } K \neq K'. %
\end{align} %
The orthogonality condition for modes with different frequency is still ambiguous. %
However, since the frequency is given as the eigenvalue of Maxwell's equation, %
the summation over $\nu$ might be similarly interpreted as that of continuous quantum states %
such as position states~\cite{Sakurai1985}. %
From the creation $(\hat{a}_K^\nu)^\dagger$ and annihilation $\hat{a}_K^\nu$ operators
in the $K$ mode with frequency $\nu$, %
we can define the number operator $\hat{n}_K^\nu \equiv (\hat{a}_K^\nu)^\dagger \hat{a}_K^\nu$ %
with a photon number state $\vert n_K^\nu \rangle$ as the corresponding eigenstate, %
such that $\hat{n}_K^\nu \vert n_K^\nu \rangle = n_K^\nu \vert n_K^\nu \rangle$. %
The general photon number state can be constructed by applying the corresponding creation operator: %
\begin{align} %
\vert n_K^\nu \rangle \equiv \frac{(\hat{a}_K^\nu)^{\dagger n}}{\sqrt{n!}} \vert 0 \rangle. %
\end{align} %
Similarly to continuous modes, such as position modes or momentum modes in quantum mechanics~\cite{Sakurai1985}, %
we can assume the completeness and orthonormality of these modes for different frequencies: %
\begin{align} %
\label{Eq:Orthonormality} %
\langle n_K^\nu \vert n_{K'}^{\nu'} \rangle = \delta_{KK'} \delta(\nu - \nu'), \\ %
\intertext{ and } %
\label{Eq:Completeness} %
\sum_\nu \sum_K \, \vert n_K^\nu \rangle \langle n_K^\nu \vert = \mathbf{1}. %
\end{align} %
According to Eqs.~\eqref{Eq:Orthonormality} and \eqref{Eq:Completeness}, %
we can treat the states over the continuous variable $\nu$ as independent to each other, %
similarly to ones over discrete variables.


To find the unknown coefficients $E_0$ and $H_0$, let's assign $\hbar \nu$ as the energy of the photonic modes: %
\begin{align} %
\frac{1}{2} \int d^3 \mathbf{r} \{ \epsilon \vert \mathbf{E}_K^\nu \vert^2 + \mu \vert \mathbf{H}_K^\nu \vert^2 \} = \hbar \nu, %
\end{align} %
where $\hbar$ is Dirac's constant. %
Since the integration along $z$-axis diverges for the propagating modes, %
we should be careful when carrying out the above integration along $z$-axis. %
Similar to the vacuum mode, we might consider a periodicity along $z$-axis, and restrict the integration range $[-L/2, L/2]$ %
for propagating modes. %
The normalization gives for a propagating TM mode ($\gamma = i \beta$) %
\begin{align} %
\label{Eq:PropagatingTMnormal} %
\vert E_0 \vert^2 = \frac{4 \hbar h_{mn}^2}{\epsilon^2 \mu \nu L S}, %
\end{align} %
for a localized TM mode ($\gamma > 0$) %
\begin{align} %
\vert E_0 \vert^2 = \frac{4\gamma \hbar h_{mn}^2}{\epsilon^2 \mu \nu S}, %
\end{align} %
for a propagating TE mode ($\gamma = i \beta$) %
\begin{align} %
\label{Eq:PropagatingTEnormal} %
\vert H_0 \vert^2 = \frac{4 \hbar h_{mn}^2}{\nu \epsilon \mu^2 L S}, %
\end{align} %
and for a localized TE mode ($\gamma > 0$) %
\begin{align} %
\vert H_0 \vert^2 = \frac{4 \gamma \hbar h_{mn}^2}{\nu \epsilon \mu^2 S}, %
\end{align} %
with the cross-sectional area of the waveguide $S= ab$. %
Since the periodicity $L$ is arbitrary, %
we need another condition for the summation over frequency for the propagating modes. %

For a description of photons in vacuum, %
the summation index over states is wave-number $k$, which satisfy periodicity $k = 2\pi n/L$ with integers $n$. %
The transition from discrete summation to continuous integration is given by~\cite{ScullZubairy1997} %
\begin{align} %
\sum_k \rightarrow \left( \frac{L}{2\pi} \right) \int dk. %
\end{align} %
Similarly, we can consider the wave-number $\beta$ along $z$-axis for the propagating modes. %
The summation over $\nu$ is divided into two parts %
by the cut-off frequency $\nu_c$ for given $K = \{ \text{TE}( \text{or } \text{TM}), m, n\}$: %
\begin{align} %
\sum_{\nu} = \sum_{\nu < \nu_c} + \sum_{\nu > \nu_c} %
= \sum_{\nu < \nu_c} + \sum_{\beta}. %
\end{align} %
Then, the summation over $\beta$ can be transformed into continuous integration similarly as is done in vacuum %
\begin{align} %
\label{Eq:ContinousBeta} %
\sum_{\beta} \rightarrow \left( \frac{L}{2\pi} \right) \int \frac{dk_z}{d\beta} d\beta %
= \left( \frac{L}{2\pi} \right) \int \frac{d\beta}{v_{gz}}, %
\end{align} %
where $v_{gz} = \frac{d\beta}{dk_z}$ is the group velocity along the $z$-axis. %
In our case, $v_{gz} = 1/\sqrt{\epsilon \mu}$, which is same to the phase velocity. %
Then the arbitrary quantization length $L$, %
shown in Eqs.~\eqref{Eq:PropagatingTMnormal} and \eqref{Eq:PropagatingTEnormal}, %
will be canceled out by the $L$ in Eq.~\eqref{Eq:ContinousBeta} in the calculation of any physical observables. %

The localized modes show exponential decay along the $z$-direction, %
and depend on the position of the two-level atom. %
However, along the $x$- and $y$-axes they show an oscillatory behavior. %
Furthermore, we emphasize again that these localized modes are also quantized modes %
since they are the eigenmodes of the two wave equations, Eq.~\eqref{Eq:WaveEqnElectric} and Eq.~\eqref{Eq:WaveEqnMagnetic}. %

\section{Spontaneous Emission} %

\subsection{Hamiltonian} %
\begin{figure}[t] %
\centering %
\includegraphics[width=0.7\linewidth]{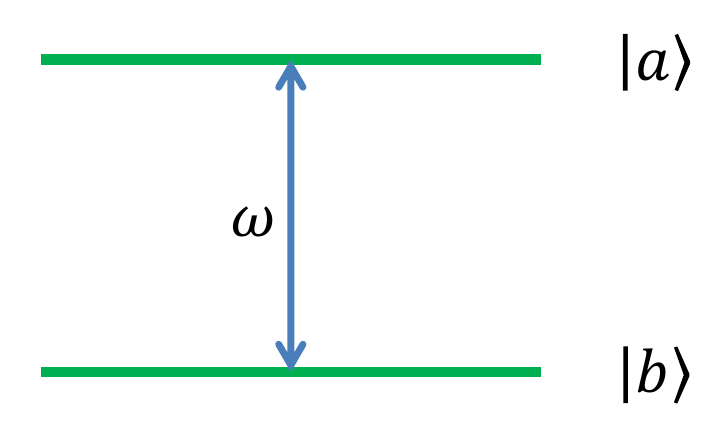} %
\caption{A two-level atom has an upper state $\vert a \rangle$ and a lower state $\vert b \rangle$. %
The transition frequency is $\omega = \omega_a - \omega_b$ with the energy of upper state $\hbar \omega_a$ %
and that of lower state $\hbar \omega_b$. (Color online)} %
\label{Fig:TwoLevelAtom} %
\end{figure}

Let's consider a spontaneous emission from a two-level atom embedded inside the PEC rectangular waveguide, %
shown in FIG.~\ref{Fig:TwoLevelAtom}. %
With the dipole approximation and rotating wave approximation, %
the Hamiltonian~\cite{ScullZubairy1997} is %
\begin{align} %
\mathcal{H} = \mathcal{H}_\text{F} + \mathcal{H}_\text{A} + \mathcal{H}_\text{int} %
\end{align} %
where %
\begin{align} %
\mathcal{H}_\text{F} & = \sum_{\nu} \, %
\sum_{K} \hbar \nu \left( (\hat{a}_K^\nu)^\dagger \hat{a}_K^\nu + \frac{1}{2} \right), \\ %
\mathcal{H}_\text{A} & = \hbar \omega_a \vert a \rangle \langle a \vert + \hbar \omega_b \vert b \rangle \langle b \vert, \\ %
\mathcal{H}_\text{int} & = - e \mathbf{r} \cdot \hat{\mathbf{E}} %
= \hbar \sum_{\nu} \, \sum_{K} g_K^\nu \hat{\sigma}_+ \hat{a}_K^\nu + \text{H.C.}. %
\end{align} %
Here, H.C. denotes Hermitian Conjugate. %
The atomic transition operators are %
\begin{align} %
\hat{\sigma}_+ & = \vert a \rangle \langle b \vert, \\ %
\hat{\sigma}_- & = (\hat{\sigma}_+)^\dagger = \vert b \rangle \langle a \vert, %
\end{align} %
and the interaction coefficient is %
\begin{align} %
g_K^\nu = - \frac{ \wp_{ba} \cdot \mathbf{E}_K^\nu }{\hbar} %
\end{align} %
with dipole transition matrix element $\wp_{ba} = e \langle b \vert \mathbf{r} \vert a \rangle$. %

\subsection{Decay rate and shift of transition energy} %

In the interaction picture, the Hamiltonian is %
\begin{align} %
\mathcal{V} = \hbar \sum_{\nu} \, \sum_{K} [ g_K^\nu \hat{\sigma}_+ \hat{a}_K^\nu e^{i(\omega - \nu) t} + \text{H.C.} ] %
\end{align} %
with $\omega = \omega_a - \omega_b$. %
Here we ignore the zero-point energy $\tfrac{1}{2} \hbar \nu$, %
since it just shifts the energy level of Hamiltonian~\cite{ScullZubairy1997}. %
From the dressed state for the two-level atom with the field %
\begin{align} %
\vert \psi(t) \rangle = c_a(t) \vert a, 0 \rangle + \sum_{\nu} \, \sum_{K} c_{b,\nu,K} \vert b, 1_K^\nu \rangle, %
\end{align} %
the corresponding Schr\"odinger equation, %
\begin{align} %
\vert \dot{\psi}(t) \rangle = - \frac{i}{\hbar} \mathcal{V} \vert \psi(t) \rangle, %
\end{align} %
gives the following equations of motion for the coefficients $c_a(t)$ and $c_{b,\nu,K}(t)$ %
\begin{align} %
\label{Eq:Eqn_ca} %
\dot{c}_a (t) & = - i \sum_{\nu} \, \sum_{K} g_K^\nu e^{i(\omega - \nu) t} c_{b,\nu,K}(t), \\ %
\label{Eq:Eqn_cbj} %
\dot{c}_{b,\nu,K} (t) & = - i (g_K^\nu)^* e^{-i(\omega - \nu) t} c_a (t), %
\end{align} %
with initial conditions $c_a(0) = 1$ and $c_{b,\nu,K}(0) = 0$, %
which means that the two-level atom is prepared in an excited state. %
Integration of Eq.~\eqref{Eq:Eqn_cbj} over $t$ and substitution into Eq.~\eqref{Eq:Eqn_ca} gives %
the following integro-differential equation for $c_a(t)$: %
\begin{align} %
\nonumber %
\dot{c}_a (t) & = - \sum_\nu \sum_{K} \vert g_K^\nu \vert^2 \int_0^t dt' e^{i(\omega - \nu) (t-t')} c_a(t') \\ %
\nonumber %
& \simeq - \sum_\nu \sum_{K} \vert g_K^\nu \vert^2 \int_0^t dt' e^{i(\omega-\nu)(t-t')} c_a(t) \\ %
\nonumber %
& = - i \sum_\nu \sum_{K} \vert g_K^\nu \vert^2 \left[ \frac{\mathcal{P}}{\omega-\nu} - i \pi \delta(\omega-\nu) \right] %
c_a(t) \\ %
& = - \left( \frac{\Gamma_\text{eff}}{2} - i \delta \omega \right) c_a(t), %
\end{align} %
where the Markovian approximation for slow varying $c_a(t)$ is used to get this term %
out of the time integration \cite{ScullZubairy1997}, %
and $\mathcal{P}$ denotes the Cauchy principal value \cite{Arken2005}. %
Here, $\Gamma_\text{eff}$ is the decay constant %
and $\delta\omega$ is the level shift of the transition from the upper level to lower one: %
\begin{align} %
\Gamma_\text{eff} & = \pi \sum_\nu \sum_K \vert g_K^\nu \vert^2 \delta(\omega - \nu), \\ %
\delta \omega & = \sum_\nu \sum_K \mathcal{P} \frac{ \vert g_K^\nu \vert^2}{\omega - \nu}. %
\end{align} %
The summation goes over all possible states including the localized modes. %

\subsection{Propagation through the waveguide} %
The temporal behavior of the excited state is given by %
\begin{align} %
c_a(t) & = e^{-(\Gamma_\text{eff}/2 - i\delta \omega) t} c_a(0) = e^{-(\Gamma_\text{eff}/2 - i\delta \omega) t} \\ %
\nonumber %
c_{b,\nu,K}(t) & = - i (g_K^\nu)^* \int_0^t dt' e^{-i(\tilde{\omega} - \nu)t' - \Gamma_\text{eff} t' /2} \\ %
& = (g_K^\nu)^* \left[ \frac{ 1 - e^{i(\nu - \tilde{\omega})t - \Gamma_\text{eff} t/2 }}{(\nu - \tilde{\omega}) + i\Gamma_\text{eff}/2} \right], %
\end{align} %
with shifted frequency $\tilde{\omega} = \omega - \delta \omega$. %
So all the quantized modes are induced by the de-excitation of the atom. %

Since we put the detector far away from the atom, the measured modes are just the propagating modes, %
because the localized modes decay exponentially along the wave-guide axis %
and can be ignored in a region near the detector. %
In addition, $c_a(t)$ is also exponentially decaying in time. %
After $t \gg \Gamma_\text{eff}^{-1}$, the dressed state becomes %
\begin{align} %
\vert \psi(t) \rangle \simeq %
\sum_{\nu} \sum_{K'} c_{b,\nu,K} \vert b, 1_K^\nu \rangle = \vert b \rangle \otimes \vert \gamma \rangle %
\end{align} %
where the photonic state far away from the atom is %
\begin{align} %
\vert \gamma \rangle = \sum_\nu \sum_{K'} %
\frac{\{ g_{K'}^\nu(\mathbf{r}_0) \}^*}{ (\nu - \tilde{\omega}) + i \Gamma_\text{eff}/2} \vert 1_{K'}^\nu \rangle. %
\end{align} %
Here, $K'$ sums over only propagating modes %
and $\mathbf{r}_0$ is the position of the atom. %

If the atomic transition frequency $\tilde{\omega}$ is between the lowest cutoff frequency and the second lowest cutoff frequency, %
the dominant propagating mode $\vert \gamma \rangle_\text{sm}$ is the TE mode with $m=1$ and $n=0$: %
\begin{align} %
\nonumber %
\vert \gamma \rangle_\text{sm} %
& = \sum_\nu (- i) \wp_{ab}^* \sqrt{ \frac{4 \nu}{\hbar \epsilon L S} } \sin \frac{\pi x_0}{a} \\ %
\label{Eq:SinglePropagationMode} %
& \qquad \times %
\frac{e^{+i\beta z_0}}{ (\nu - \tilde{\omega}) + i \Gamma_\text{eff}/2} \vert 1_{\text{TE},1,0}^\nu \rangle, %
\end{align} %
where the atom's position $\mathbf{r}_0 = (x_0, y_0 , z_0)$. %
It is obvious that the generated propagating mode $\vert \gamma \rangle_\text{sm}$ %
is independent of atom's position along the $y$-axis. %


If the frequency $\omega$ is below the lowest cutoff frequency, %
the induced modes from the excited atom are localized modes, %
and the energy will oscillate between the atom and the localized modes. %

\subsection{Measurement by the detector} %

A single photon can be measured by a detector through photon absorption process, or by another atom through excitation. %
Here, we consider a measurement by a detector, %
which can be described by the correlation function \cite{ScullZubairy1997}: %
\begin{align} %
\nonumber %
G^{(1)} (\mathbf{r},t) & = \langle \psi \vert \hat{\mathbf{E}}^{(-)}(\mathbf{r},t) \hat{\mathbf{E}}^{(+)}(\mathbf{r},t) %
\vert \psi \rangle \\ %
& = \vert \langle 0 \vert \hat{\mathbf{E}}^{(+)}(\mathbf{r},t) \vert \gamma \rangle \vert^2, %
\end{align} %
where the positive frequency part in electric field operator from Eq.~\eqref{Eq:EFieldOperator} is %
\begin{align} %
\label{Eq:Positive_EFieldOperator} %
\hat{E}^{(+)}(\mathbf{r},t) = \sum_\nu %
\sum_K \mathbf{E}_K \hat{a}_K e^{-i\nu t}. %
\end{align} %

For the single photonic mode case, Eq.~\eqref{Eq:SinglePropagationMode}, %
we obtain the following expression through a contour integral, which is explicitly derived in the Appendix: 
\begin{align} %
\nonumber %
\langle 0 \vert \hat{\mathbf{E}}^{(+)}(\mathbf{r},t) \vert \gamma \rangle_\text{sm} %
= & i \frac{\tilde{\omega} \sqrt{\epsilon \mu} \wp_{ab}^*}{\pi^2 \epsilon S}
\frac{\tilde{\omega}}{\sqrt{(\pi/a)^2 - \tilde{\omega}^2}} \\ %
\nonumber %
& \quad \times \sin \frac{\pi x_0}{a} \sin \frac{\pi x}{a} %
\Theta(t - \sqrt{\epsilon \mu} z) \\ %
\label{Eq:ContourElectric} %
& \quad \times e^{i (\beta_r - i \beta_i) \Delta z - i (\tilde{\omega} - i \Gamma_\text{eff}/2) t}, %
\end{align} %
where the distance $\Delta z$ is from the source atom to the field operator evaluation point along $z$-direction, %
and $\Theta$ is the step function. %
Explicit expressions for $\beta_r$ and $\beta_i$
are shown in the Appendix. 
One thing to note is that the above equation is evaluated %
at the shifted resonance frequency $\tilde{\omega} = \omega - \delta \omega$. %

\begin{figure}[ht] %
\centering %
\includegraphics[width=1.00\linewidth]{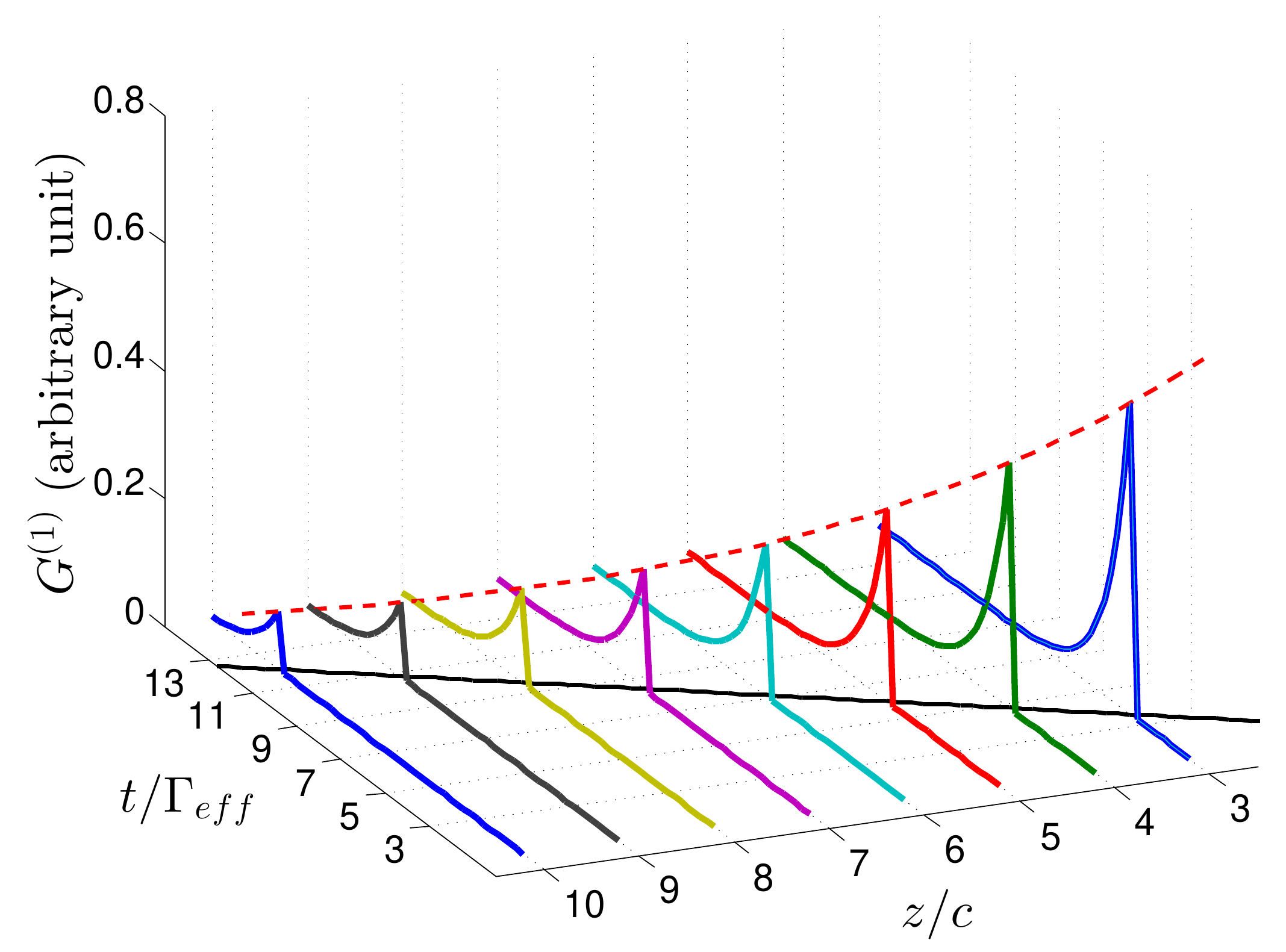} %
\caption{Profile of correlation function in $(z,t)$ plane. %
The distant detector at $z$ can detect after some time $t$ allowable by the causality relation, %
shown as black solid line on $(z,t)$ plane. %
In a time $t$ along $z$ axis, the detection probability shows exponential behavior with decay constant $\Gamma_\text{spa}$, %
drawn by red dashed line. %
Similarly, the detection probability at fixed position is maximum at the initial time and shows exponential decay along $t$ %
with decay constant rate $\Gamma_\text{eff}$. %
Generally, $\Gamma_\text{eff}$ and $\Gamma_\text{spa}$ are different. %
The maximum value of correlation at $z$ shows also the exponential behavior. %
The values of parameters are $\sqrt{\epsilon\mu} = 1.2$ and $\Gamma_\text{spa} / \Gamma_\text{eff} = 0.8$. %
(Color online)} %
\label{Fig:ProfileCorrelation} %
\end{figure} %

Then, the first-order correlation function is %
\begin{widetext} %
\begin{align} %
\label{Eq:1stCorrelation} %
G^{(1)}_\text{sm} ( \mathbf{r},t) = %
\left( \frac{\tilde{\omega} \sqrt{\epsilon \mu} \wp_{ab}^*}{\pi^2 \epsilon S} \right)^2 %
\frac{\tilde{\omega}^2}{(\pi/a)^2 - \tilde{\omega}^2} %
\sin^2 \frac{\pi x}{a} \sin^2 \frac{\pi x_0}{a} %
\Theta( t - \sqrt{\epsilon \mu} \Delta z) %
e^{ \Gamma_\text{spa} \Delta z - \Gamma_\text{eff} t}, %
\end{align} %
\end{widetext} %
where the spatial decay rate is %
\begin{align} %
\Gamma_\text{spa} = 2\beta_i. %
\end{align} %
One thing to note is that the ratio between the spatial decay rate $\Gamma_\text{spa}$ %
and the temporal decay rate $\Gamma_\text{eff}$ %
is generally different from the velocity $\sqrt{\epsilon\mu}$. %

In Fig.~\ref{Fig:ProfileCorrelation}, we show the profile of the correlation function in $(z,t)$ plane. %
The distant detector at $z$ from the source atom, located at $z_0$, can detect %
after a propagation time $t = \sqrt{\epsilon \mu} \vert z - z_0 \vert$ by the causality relation. %
Along the $z$ axis, the detection probability shows exponential behavior with decay constant $\Gamma_\text{spa}$. %
Similarly, the detection probability at a fixed position is maximum at the arrival of the photon %
and shows exponential decay along $t$ with decay constant rate $\Gamma_\text{eff}$. %
It also shows the exponential behavior along the line with $t-\sqrt{\epsilon \mu} z = \text{const.}$, %
since the ratio is different from $\sqrt{\epsilon \mu}$. %


In Eq.~\eqref{Eq:1stCorrelation}, %
the spatial decay and temporal decay are different in the first order correlation function %
due to the geometrical boundary of the waveguide. %
The ratio of two decay rates $\Gamma_\text{spa}/\Gamma_\text{eff}$ is equal to $\sqrt{\epsilon \mu}$
at $\tilde{\omega} = \tilde{\omega}_d$ with %
\begin{align} %
\tilde{\omega}_d^2 = \frac{ \epsilon^2 \mu^2 \Gamma_\text{eff}^4 + 12 \epsilon \mu \Gamma_\text{eff}^2 (\pi/a)^2 + 4 (\pi/a)^4} %
{8 \{ \epsilon \mu \Gamma_\text{eff}^2 + 2 (\pi/a)^2 \} \epsilon \mu}, %
\end{align} %
which is derived from Eq.~\eqref{Eq:beta_i} in the Appendix. 
In the limit $\omega \rightarrow \infty$, %
the spatial decay rate approaches the temporal decay rate. %
In the $a \gg b$ limit (that is, like a shallow slab), %
the temporal decay rate is larger than the spatial one when $\tilde{\omega}^2 > \Gamma_\text{eff}^2/8$. %
The correlation function is maximum in the cross-section of the waveguide %
when the detector and the embedded atom are in the middle of the waveguide along the $x$ axis. %

In comparison, the correlation function in free space \cite{ScullZubairy1997} is %
\begin{align} %
G^{(1)}_\text{free} (\mathbf{r},t) = \frac{ \vert \mathcal{E}_0 \vert^2}{\vert \mathbf{r} - \mathbf{r}_0 \vert^2} %
\Theta \left( t - \frac{ \vert \mathbf{r} - \mathbf{r}_0 \vert}{c} \right) %
e^{ - \Gamma ( t - \vert \mathbf{r} - \mathbf{r}_0 \vert / c) }, %
\end{align} %
where $c$ is the speed of light, %
and the vacuum decay rate %
\begin{align} %
\Gamma = \tfrac{1}{4\pi\epsilon_0} \tfrac{4\omega^3 \wp_{ab}^2}{3\hbar c^3} %
\end{align} %
with the dipole moment $\wp_{ab}$ \cite{ScullZubairy1997}. %
Here, %
\begin{align} %
\mathcal{E}_0 = - \tfrac{\omega^2 \wp_{ab} \sin \eta}{4\pi \epsilon_0 c^2 \Delta r}, %
\end{align} %
and $\eta$ is the angle of the dipole moment from $z$-axis. %

\section{Conclusion} %

A simple model for the spontaneous emission problem is reconsidered in terms of the second quantization formalism. %
It has been revealed that the usually ignored localized modes should be considered %
as one of the quantized modes to calculate the decay rate for spontaneous emission. %
However, the description for propagation and detection of photon is similar to classical treatment %
since localized modes cannot contribute to these phenomena. %
One thing to notice is that the decay rates in space and time are different with smaller spatial decay rate usually. %
The problem with dispersive dielectric material will be investigated later. %

Furthermore, since we get the quantized electric field operator in this simple system, %
we can describe any quantum optical phenomena also. %
However, to extend Glauber's method \cite{Glauber1991} to other systems might be difficult %
since it is usually difficult to calculate the complete set of eigenfunctions. %
To overcome this obstacle, approximate methods to the quantized field should be developed, %
such as to numerically evaluate the eigenmodes for a description of the behavior of the optical system. %

\appendix

\section*{Appendix: Evaluation of the first-order Correlation function} %
\label{Appendix:Cor} %

The calculation for the transition probability amplitude at the detection from a propagating photonic mode to vacuum mode %
is explicitly shown. %
From Eqs.~\eqref{Eq:SinglePropagationMode} and \eqref{Eq:Positive_EFieldOperator}, %
\begin{align} %
\nonumber %
& \langle 0 \vert \hat{\mathbf{E}}^{(+)}(\mathbf{r},t) \vert \gamma \rangle_\text{sm} \\ %
\nonumber %
= & \sum_\nu \mathbf{E}_{\text{TE},10} e^{-i\nu t} %
\frac{g_{\text{TE},10}^*(r_0)}{(\nu - \tilde{\omega}) + i \Gamma_\text{eff}/2} \\ %
%
\nonumber %
\simeq & - \frac{2\sqrt{\epsilon \mu}}{\pi \epsilon S}
\sin \frac{\pi x_0}{a} \sin \frac{\pi x}{a} %
\int d\beta \frac{\nu \wp_{ab}^*  e^{-i\beta(z- z_0) - i\nu t}}{(\nu - \tilde{\omega}) + i \Gamma_\text{eff}/2} \\ %
\label{Eq:Contour_Init} %
\simeq & - \frac{2\tilde{\omega} \sqrt{\epsilon \mu} \wp_{ab}^*}{\pi \epsilon S}
\sin \frac{\pi x_0}{a} \sin \frac{\pi x}{a} %
\int d\beta \frac{e^{-i\beta(z- z_0) - i\nu t}}{(\nu - \tilde{\omega}) + i \Gamma_\text{eff}/2} %
\end{align} %
where the frequency $\nu$ is approximated by the shifted resonance frequency $\tilde{\omega}$.

\begin{figure}[ht] %
\centering %
\includegraphics[width=0.80\linewidth]{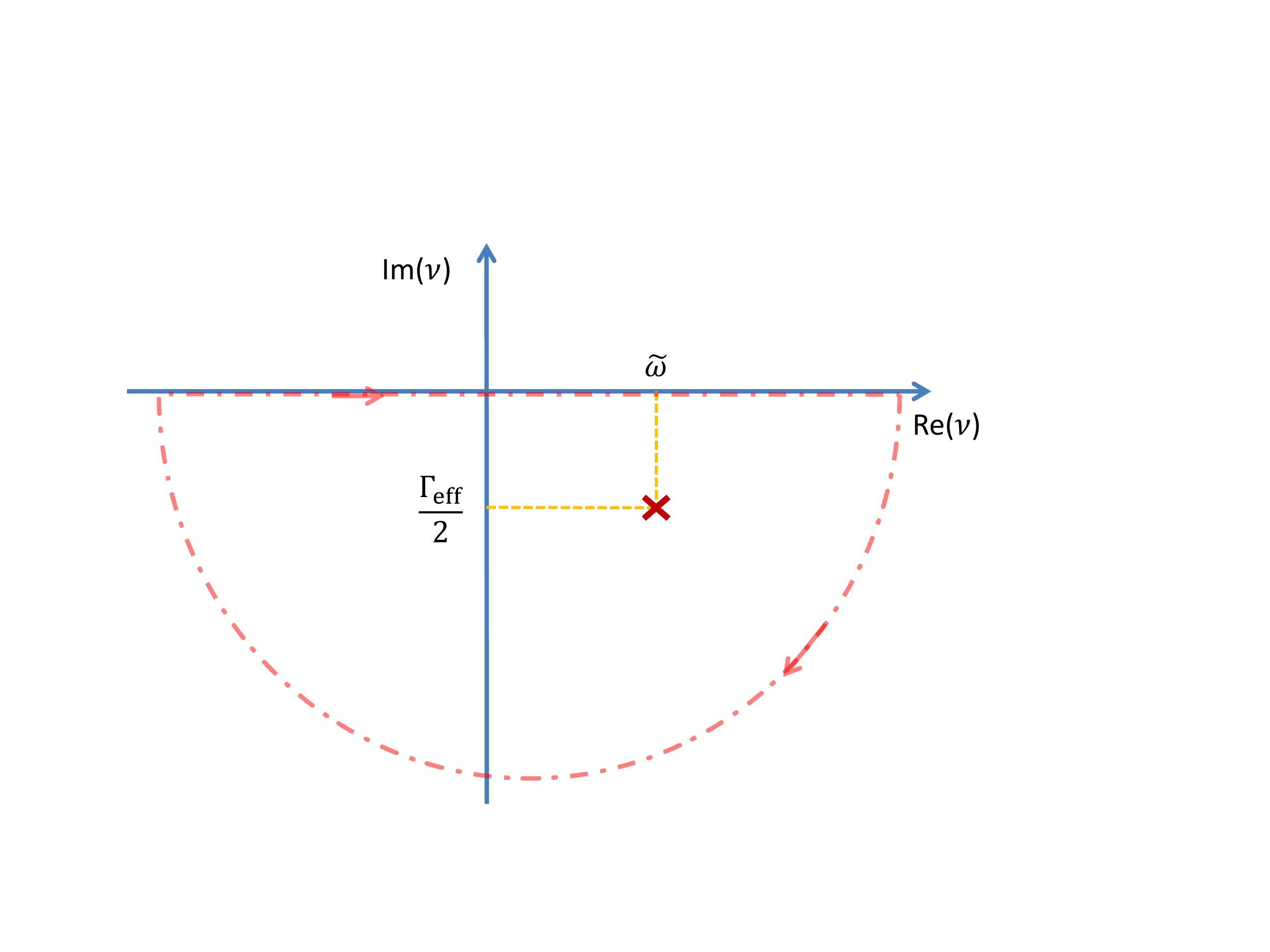} %
\caption{Profile of contour integral in complex $\nu$-plane. (Color online)} %
\label{Fig:ContourIntegral} %
\end{figure} %

To evaluate the above integration in Eq.~\eqref{Eq:Contour_Init}, %
let's consider the contour integration after change of variables from $\beta$ to $\nu$. %
Since we are interested in a region near $\tilde{\omega}$ where the pole is located, %
\begin{align} %
\nonumber %
& \int d\beta \frac{e^{i\beta ( z_0 - z) - i \nu t}}{(\nu - \tilde{\omega}) + i \Gamma_\text{eff}/2} \\ %
\nonumber %
\simeq & \sqrt{\epsilon\mu} \int \frac{\nu}{\beta} d\nu \frac{e^{i\beta ( z_0 - z) - i \nu t}}{(\nu - \tilde{\omega}) + i \Gamma_\text{eff}/2} \\ %
\simeq & \sqrt{\epsilon\mu} \frac{\tilde{\omega}}{\sqrt{(\pi/a)^2 - \tilde{\omega}^2}} \int d\nu \frac{e^{i\beta ( z_0 - z) - i \nu t}}{(\nu - \tilde{\omega}) + i \Gamma_\text{eff}/2}.
\end{align} %
Since the pole is at lower-half plane in complex $\nu$-plane, %
the contour $\mathcal{C}$ is as shown in Fig.~\ref{Fig:ContourIntegral} to get the nonzero result. %
\begin{align} %
\nonumber %
& \int d\nu \frac{e^{i\beta ( z_0 - z) - i \nu t}}{(\nu - \tilde{\omega}) + i \Gamma_\text{eff}/2} \\ %
\nonumber %
= & \frac{1}{2\pi i} %
\int_\mathcal{C} d\nu \frac{e^{i\beta ( z_0 - z) - i \nu t}}{(\nu - \tilde{\omega}) + i \Gamma_\text{eff}/2} \\ %
\label{Eq:ContourIntegral_result} %
= & \frac{1}{2\pi i} \Theta(t - \sqrt{\epsilon \mu} z) %
e^{i (\beta_r - i \beta_i) \Delta z - i (\tilde{\omega} - i \Gamma_\text{eff}/2) t}, %
\end{align} %
where $\Delta z = \vert z - z_0 \vert$ is the distance between the source atom and a detector, %
and $\beta_r$ and $\beta_i$ are %
\begin{align} %
\nonumber %
\beta_r + i \beta_i & = \beta(\nu = \tilde{\omega} - i \Gamma_\text{eff}/2) \\ %
\nonumber %
& = \sqrt{ \sqrt{\epsilon \mu} \left( \tilde{\omega} - i \frac{\Gamma_\text{eff}}{2} \right)^2 %
- \left( \frac{\pi}{a} \right)^2 } \\ %
& = \sqrt{ \sqrt{\epsilon \mu} ( \tilde{\omega}^2 - \frac{\Gamma_\text{eff}^2}{4} ) - \left( \frac{\pi}{a} \right)^2 %
- i \sqrt{\epsilon \mu} \tilde{\omega} \Gamma_\text{eff} } \, . %
\end{align} %
After a simple evaluation, we get %
\begin{align} %
\label{Eq:beta_r} %
\beta_r & = \sqrt{ \frac{ A + \sqrt{ A^2 + 4 B^2}}{2}} > 0, \\ %
\label{Eq:beta_i} %
\beta_i & = - \sqrt{ \frac{ \sqrt{ A^2 + 4 B^2} - A}{2}} %
\end{align} %
with $A = \sqrt{\epsilon \mu} \left( \tilde{\omega}^2 - \Gamma_\text{eff}^2/4 \right) - \pi^2/a^2$, %
and $B = \sqrt{\epsilon \mu}\tilde{\omega} \Gamma_\text{eff}/2$. %
Here, we choose $\beta_r$ as positive. %

Substituting Eq.~\eqref{Eq:ContourIntegral_result} into Eq.~\eqref{Eq:Contour_Init} we obtain %
the transition probability amplitude %
\begin{align} %
\nonumber %
\langle 0 \vert \hat{\mathbf{E}}^{(+)}(\mathbf{r},t) \vert \gamma \rangle_\text{sm} %
= & i \frac{\tilde{\omega} \sqrt{\epsilon \mu} \wp_{ab}^*}{\pi^2 \epsilon S}
\frac{\tilde{\omega}}{\sqrt{(\pi/a)^2 - \tilde{\omega}^2}} \\ %
\nonumber %
& \quad \times %
\sin \frac{\pi x_0}{a} \sin \frac{\pi x}{a} %
\Theta(t - \sqrt{\epsilon \mu} z) \\ %
& \quad \times %
e^{i (\beta_r - i \beta_i) \Delta z - i (\tilde{\omega} - i \Gamma_\text{eff}/2) t}. %
\end{align} %

\begin{acknowledgments} %
This work was supported by %
the Air-Force office of Scientific Research and the National Science Foundation. %
\end{acknowledgments} %

\end{document}